%% file: main.tex
\documentclass[conference]{IEEEtran}
\IEEEoverridecommandlockouts
\usepackage{cite}
\usepackage{amsmath,amssymb,amsfonts}
\usepackage{algorithmic}
\usepackage{graphicx}
\usepackage{textcomp}
\usepackage{xcolor}
\usepackage{optidef}
\usepackage[commentmarkup=todo,todonotes={textsize=tiny,textwidth=1.3cm}]{changes}
\definechangesauthor[name=Martin, color=green]{MK}
\definechangesauthor[name=Peter, color=blue]{PZ}

\usepackage{amsmath}
\usepackage{pgfplots}
\usepackage{pgfplotstable}
\pgfplotsset{compat=1.14}
\usepackage{placeins}
\usepackage{pdflscape}
\usepackage{adjustbox}
\usepackage{tkz-fct}
\usetikzlibrary{arrows}
\usetikzlibrary{circuits}
\usetikzlibrary{decorations}
\usepackage{hhline,multirow}
\usepackage{rotating}
\usepackage{booktabs}

\newcolumntype{C}[1]{>{\centering\arraybackslash}p{#1}} 
\newcolumntype{R}[1]{>{\raggedleft\arraybackslash}p{#1}} 
\newcolumntype{L}[1]{>{\raggedright\arraybackslash}p{#1}} 

\tikzset{
	png export/.style={
		external/system call/.add=
		{}
		{; convert -density 300 -transparent white "\image.pdf" -quality 100 "\image.png";},
		/pgf/images/external info,
		/pgf/images/include external/.code={%
			\includegraphics
			[width=\pgfexternalwidth,height=\pgfexternalheight]
			{##1.png}%
		}
	}
}
\tikzset{
	eps export/.style={
		external/system call/.add=
		{}
		{; convert -density 300 "\image.pdf" -quality 100 "\image.eps";},
		/pgf/images/external info,
		/pgf/images/include external/.code={%
			\includegraphics
			[width=\pgfexternalwidth,height=\pgfexternalheight]
			{##1.eps}%
		}
	}
}
\IEEEoverridecommandlockouts
\IEEEpubid{\makebox[\columnwidth]{978-1-5386-8398-9/18\$31.00 \copyright 2019 IEEE \hfill} \hspace{\columnsep}\makebox[\columnwidth]{ }}

\def\BibTeX{{\rm B\kern-.05em{\sc i\kern-.025em b}\kern-.08em
    T\kern-.1667em\lower.7ex\hbox{E}\kern-.125emX}}
\begin{document}

\title{Efficient Error-Tolerant\\ Quantized Neural Network Accelerators}

\author{\IEEEauthorblockN{Giulio Gambardella\IEEEauthorrefmark{1}, Johannes Kappauf\IEEEauthorrefmark{1}, Michaela Blott\IEEEauthorrefmark{1}, 
Christoph Doehring\IEEEauthorrefmark{2}\\Martin Kumm\IEEEauthorrefmark{3}, Peter Zipf\IEEEauthorrefmark{4} and Kees Vissers\IEEEauthorrefmark{1}} \IEEEauthorblockA{\IEEEauthorrefmark{1}Xilinx Research Labs, Dublin, Ireland, Email: \{giuliog,johannes,mblott,kees.vissers\}@xilinx.com} \IEEEauthorblockA{\IEEEauthorrefmark{2}
Coburg University, Germany, Email: 
doch1501@stud.hs-coburg.de
} \IEEEauthorblockA{\IEEEauthorrefmark{3}
Fulda University of Applied Sciences, Germany, Email: martin.kumm@cs.hs-fulda.de
}
\IEEEauthorblockA{\IEEEauthorrefmark{4}
University of Kassel, Digital Technology Group, Kassel, Germany, Email: zipf@uni-kassel.de
}
}

\maketitle

\begin{abstract}

Neural Networks are currently one of the most widely deployed machine learning algorithms. In particular, Convolutional Neural Networks (CNNs), are gaining popularity and are evaluated for deployment in safety critical applications such as self driving vehicles.
Modern CNNs feature enormous memory bandwidth and high computational needs, challenging existing hardware platforms to meet throughput, latency and power requirements. Functional safety and error tolerance need to be considered as additional requirement in safety critical systems.
In general, fault tolerant operation can be achieved by adding redundancy to the system, which is further exacerbating the computational demands. 
Furthermore, the question arises whether pruning and quantization methods for performance scaling turn out to be counterproductive with regards to fail safety requirements.  
In this work we present a methodology to evaluate the impact of permanent faults affecting Quantized Neural Networks (QNNs) and how to effectively decrease their effects in hardware accelerators. 
We use FPGA-based hardware accelerated error injection, in order to enable the fast evaluation. 
A detailed analysis is presented 
showing that QNNs containing convolutional layers are by far not as robust to faults as commonly believed and can lead to accuracy drops of up to 10\%.
To circumvent that, we propose two different methods to increase their robustness: 1) selective channel replication 
which adds significantly less redundancy than used by the common triple modular redundancy
and 2) a fault-aware scheduling of processing elements for folded implementations.
\end{abstract}

\begin{IEEEkeywords}
neural networks, safety, automotive, FPGA, quantized neural networks
\end{IEEEkeywords}

\input{introduction.tex}

\input{background.tex}

\input{campaign.tex}

\input{hardware-independent.tex}
\input{hardware-dependent.tex}

\input{conclusions.tex}

\bibliographystyle{ieee/IEEEtran} 
\bibliography{ieee/IEEEexample}

\end{document}

%% file: introduction.tex
\section{Introduction}
\label{sec:introduction}

In order to use an electronic device in a safety critical
application, its dependability must be evaluated, usually composed of reliability, availability, maintainability and safety (RAMS). 
The functional safety analysis and its evaluation is regulated by
standards depending on the application domain (e.g., IEC-61508
for industrial systems, ISO-26262 for road vehicles and EN 50126/8/9 for rail transport), 
with safety levels proportional to the criticality of the application.
Failure modes, effects, and diagnostic analysis (FMEDA)~\cite{FMEDA} have to be evaluated for each of the components of a design in order to model the system's safety features.

Within machine learning, deep learning and especially CNNs have recently gained major visibility due to their high accuracy in many computer vision tasks. Nonetheless, their algorithm complexity is associated with enormous compute and memory requirements.
Significant efforts have been made, tackling the inherent redundancy of neural networks by means of weights and synapse pruning, using less intensive layers (depth-wise separable convolution~\cite{MobileNet}) and non-arithmetic layers (e.g., ShiftNet~\cite{Shiftnet}) or applying parameter quantization~\cite{binary_net}. 
Many of those techniques require ad-hoc hardware back-ends to fully exploit the optimization strategies, making programmable devices like Field Programmable Gate Arrays (FPGAs) and Adaptive Compute Acceleration Platforms (ACAPs) very interesting implementation targets due to their reconfigurability and flexibility. 
Additionally, the programmability of FPGAs enables hardware architecture changes to customize the application demands, making those devices appealing for safe implementations of hardware accelerators.
The need for dependable electronic systems in safety critical applications has led to the need for fast, reliable and affordable methodologies to assess and measure safety.
In this paper we present hardware accelerated error injection for quantized neural networks. The main contributions of this paper are:
\begin{itemize}
    \item High confidence, bit-accurate and high-speed error injection by means of FPGA implementations;
    \item Two orthogonal methodologies for deriving optimized hardware implementations for guaranteed worst-case accuracy drops in case of single errors.
\end{itemize}

\section{Related Work}
\label{sec:related}

Multiple recent works are trying to assess the safety of neural networks, especially targeting autonomous driving systems~\cite{art2018_nvidia}.
To evaluate the safety level, fault or error injection can be used to get precise diagnostic coverage data but requires much more power in terms of computation, usually resorting to software simulation or hardware emulation.
In order to overcome this problem, Bosio et al.~\cite{art2018_bosio} proposed the adoption of statistical permanent fault injection to decrease the computational needs, at the expense of confidence in the measured diagnostic coverage. 
Gehr et al.~\cite{Ai2_ETH} rely on abstract interpretation to model neural network layers, in order to perform error injection at higher abstraction levels, thus decreasing the simulation time. 
In~\cite{Ares_DAC}, a framework for bit-level error injection in the memory subsystem in Keras is presented, showing a 12\% difference in resilience analysis with respect to an hardware implementation for a multi-layer perceptron. 
Multiple works are targeting soft-errors only, either by performing software emulation~\cite{Understanding_error_propagation} or by mean of neutron beam testing~\cite{Rech_2017},
proving some inherent resilience of neural networks. 
Even when heavily quantized, CNNs still provide some built-in fault tolerance, as shown by Nunez-Yanez et al.~\cite{Nunez-TOC} while the system is stressed under Dynamic Voltage Scaling (DVS) and Dynamic Frequency Scaling (DFS). 


%% file: background.tex
\section{Background}
\label{sec:background}

\subsection{Neural Network Acceleration and Scheduling}

CNNs are usually composed of a sequence of layers, each with its characteristics in terms of feature map sizes, channels and filters. We refer the readers to~\cite{Guo2017,Mittal2018,vkb18} for an exhaustive list of CNN accelerators targeting programmable logic. In this paper, we adopted the QNNs accelerator, code-named FINN, and described in~\cite{finn,finn-r} which are publicly available at~\cite{BNN-PYNQ-REPO}.
CNN hardware accelerators consist in general of an array of Processing Elements (PEs) on which layers or portion of layers are scheduled to be executed in sequence, with the network parameters either residing on-chip or in external memory.
Whenever a single PE is faulty, it is thus going to affect multiple outputs both within the layer or among layers. The portion of the network affected by the corrupted PE depends on the scheduling of the neural network. 
For FINN, due to the data flow structure instantiating one compute block per layer (with multiple PEs per compute block) a single faulty PE will affect results of a single layer. 
This is visualized in Figure~\ref{fig:folding}, which shows a case in which 3 PEs are instantiated to compute 6 output channels of a layer%
. The ratio between channels and PEs gives the number of clock cycles required for computing an output pixel and is defined as the folding factor $f$, in this case $f=6/3=2$.

\input{images/folding.tex}

\subsection{Error Injection Model}

In general, QNNs are using potentially different precisions for their weights W and activations A, denoted in the following as W$w$A$a$, where $w$ and $a$ are the respective bit widths. Depending on the activation precision, neurons can attain different activation stages which we are injecting. The evaluated networks are thereby all using symmetric activations around zero. Thus, assuming a ternary activations network (i.e., $a=2$), activations can have a value out of \{-1, 0, 1\}, which is equal to the possible stuck-at errors in our error injection context using this precision. When setting neurons in the network to stuck-at values, we are thus always injecting $a$ bits at once.



%% file: images/folding.tex
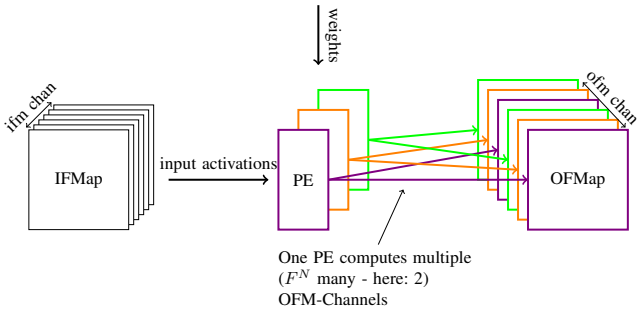
\begin{figure}
\centering
\resizebox{\columnwidth}{!}{
\begin{tikzpicture}
	\draw[->, line width=0.4mm] (2.8,1) -- (4.8,1) node[pos=0.5,sloped,above] {input activations};
	\draw[->, line width=0.4mm] (5.8,4.5) -- (5.8,3.3) node[pos=0.5,sloped,above] {weights};
	\foreach \a in {5,...,0}
		\def \x { \a / 10}
		\filldraw[fill=white, draw=black] (0 + \x,0 + \x) rectangle (2 + \x,2 + \x);	
	\def \a {4}
	\def \x { \a / 5}
	\filldraw[fill=white, draw=green, line width = 0.4mm] (5 + \x,0 + \x) rectangle (6 + \x,2 + \x);	
	\def \a {2}
	\def \x { \a / 5}
	\filldraw[fill=white, draw=orange, line width = 0.4mm] (5 + \x,0 + \x) rectangle (6 + \x,2 + \x);	
	\def \a {0}
	\def \x { \a / 5}
	\filldraw[fill=white, draw=violet, line width = 0.4mm] (5 + \x,0 + \x) rectangle (6 + \x,2 + \x);

	\def \a {5}
	\def \x { \a / 5}
	\filldraw[fill=white, draw=green, line width = 0.4mm] (10 - \x,0 + \x) rectangle (12 - \x,2 + \x);		
	\def \a {4}
	\def \x { \a / 5}
	\filldraw[fill=white, draw=orange, line width = 0.4mm] (10 - \x,0 + \x) rectangle (12 - \x,2 + \x);		
	\def \a {3}
	\def \x { \a / 5}
	\filldraw[fill=white, draw=violet, line width = 0.4mm] (10 - \x,0 + \x) rectangle (12 - \x,2 + \x);	
	\def \a {2}
	\def \x { \a / 5}
	\filldraw[fill=white, draw=green, line width = 0.4mm] (10 - \x,0 + \x) rectangle (12 - \x,2 + \x);		
	\def \a {1}
	\def \x { \a / 5}
	\filldraw[fill=white, draw=orange, line width = 0.4mm] (10 - \x,0 + \x) rectangle (12 - \x,2 + \x);		
	\def \a {0}
	\def \x { \a / 5}
	\filldraw[fill=white, draw=violet, line width = 0.4mm] (10 - \x,0 + \x) rectangle (12 - \x,2 + \x);		
	
	\draw[->, color=violet, line width = 0.4mm] (6, 1) -- (10, 1);	
	\draw[->, color=violet, line width = 0.4mm] (6, 1) -- (10 - 0.6, 1 + 0.6);	
	\draw[->, color=orange, line width = 0.4mm] (6 + 0.4, 1 + 0.4) -- (10 - 0.2, 1 + 0.2);	
	\draw[->, color=orange, line width = 0.4mm] (6 + 0.4, 1 + 0.4) -- (10 - 0.8, 1 + 0.8);		
	\draw[->, color=green, line width = 0.4mm] (6.8, 1.8) -- (10 - 0.4, 1 + 0.4);	
	\draw[->, color=green, line width = 0.4mm] (6.8, 1.8) -- (10 - 1.0, 1 + 1.0);	
	\draw[<->] (-0.05, 2.05) -- (0.45,2.55) node[pos=0.5,sloped,above] {ifm chan};
	\draw[<->] (11.95, 2.05) -- (11.1,2.95) node[pos=0.5,sloped,above] {ofm chan};
	\draw(1,1) node {IFMap};	
	\draw(5.5,1) node {PE};	
	\draw(11,1) node {OFMap};	
	\draw[->] (7, -0.3) -- (7.5, 0.8) node[pos = 0.0, below, text width = 4cm] {One PE computes multiple ($F^N$ many - here: 2) OFM-Channels};	
\end{tikzpicture}
}
\caption{Output channels folding in FINN} \label{fig:folding}
\vspace{-0.5cm}
\end{figure}

%% file: campaign.tex
\section{Error Injection Methodology}
\label{sec:errorinjection}
The adopted implementation of the FINN~\cite{finn} compute data path, as an example of a typical CNN compute fabric, is based on data flow implementations of all the neural network layers. Each layer is built upon a Matrix-Vector Threshold Unit (MVTU) performing the computation. The MVTU consists of an array of multiply and accumulate engines performing the matrix-vector multiplication, and a thresholding block. This is shown in Figure~\ref{fig:satchannel}.
The thresholding block performs a comparison between the accumulation results and a set of thresholds, which are computed at design time by fusing biases, batch-normalization and activation quantization~\cite{finn}. Since batch-normalization and biases, as resulted from training, are different for each OFM channel, a different set of thresholds has to be computed for each channel.  
In detail, the activation within an MVTU is performed by accumulating the results of $\sum_{i=1}^N (val > th_i$), 
where $val$ is the result of the matrix-vector accumulation operation and $th_i$ is each of the computed thresholds. 
Assuming a fully binarized network (W1A1), it is possible to inject stuck-at 1 (s@1) and stuck-at -1 (s@-1) errors by setting those threshold values during run-time. 
We refer as s@-1 for BNN since -1 is the logical value of 0 in a BNN.
A s@-1 can be implemented by fixing the $th$ to $th_\text{max}$ where $th_\text{max}$ is a value which is bigger than any input can reach. Thus, the activation is never performed.
Similarly, s@1 can be implemented by fixing the $th$ to $-th_\text{max}$. The accumulation will always be greater than the regarding threshold. In case of multiple bit output, it is possible to fix the result activation to any possible value by fixing as many thresholds as needed to $-th_\text{max}$ and the others to $th_\text{max}$.

In all of the following analysis experiments, we decided to use the overall accuracy of the neural network as a figure of merit, measured by running inference on the complete testset (10,000 images, as used during training to validate accuracy) in the error-injected system. We are evaluating two different topologies, namely CNV (trained on CIFAR-10) and LFC (trained on MNIST), as introduced in FINN \cite{finn}. The networks are implemented with different precisions for weights activations to analyze accuracy and error tolerance with the precision.
LFC is a multi-layer perceptron (MLP) with 3 fully connected layers while CNV is a CNN with 6 convolutional layers, 3 max pooling layers and 3 fully connected layers. 

As shown in \cite{Fault_Models_for_Neural_Hardware}, neuron error (stuck-at) is one of the common error models for feed-forward artificial neural networks. This model perfectly applies to fully connected layers, in which a single error will effect a single output pixel, being the output a single dimensional feature map.
Expanding on the concept to CNNs, a neuron stuck-at will affect a complete output channel of a layer. 
We refer to error injection instead of fault injection since we fix the activations of a whole output channel to one of the possible activations values. This means that, for multi-bit activations, multiple faults are actually injected.  


%% file: hardware-independent.tex
\section{Generic Model Optimization}
\label{sec:hardwareindependent}
\subsection{Analysis of Stuck-At Errors}

The goal of the generic model analysis is to evaluate the tolerance of the neural network itself, without considering the actual specifics of a given hardware architecture and the associated compute schedule of the layers and channels. 


As explained in Section~\ref{sec:errorinjection}, we run experiments and evaluate accuracy when a complete output feature map (OFMap) channel (ofm chan) of a layer is stuck-at a fixed value, as shown in Figure~\ref{fig:satchannel}. 
For each output channel a set of thresholds is evaluated. In order to perform error injection on a complete channel, it is sufficient to fix the related threshold values as explained in Section~\ref{sec:errorinjection}. 

\input{images/satchannel.tex}

Tables~\ref{tab:satchannel-lfc-summary} and \ref{tab:satchannel-cnv-summary} list the results of the injection campaign performed on LFC and CNV, respectively, at multiple precisions. For each network precision, we list the minimum accuracy achieved with one of the channel stuck-at experiments, as well as the maximum among all channels.
For the fully binarized neural network, as shown in the first row of  both tables, we report results of injection of s@$-1$ and s@1 having a binary representation of 0 and 1, respectively.
When a higher bitwidth is used for activations, as in W1A2 and W2A2, we perform injection for each of the possible values of the activation. In the 2-bit activation case (actually ternary due to symmetric quantization) we inject the values $-1$, 0 and +1, having a binary representations of 11, 00 and 01, respectively.

From the LFC results in Table~\ref{tab:satchannel-lfc-summary}, the overall accuracy is only slightly impacted, for all tested precisions, showing a worst-case accuracy drop of only 0.09 \% in the W2A2 case for s@$-1$. 
This is mainly due to the fact that only fully connected layers are used which are inherently very redundant and able to easily tolerate a single neuron generating error results. 

The situation changes significantly when the network contains convolutional layers like for CNV as shown in Table~\ref{tab:satchannel-cnv-summary}.
For the W1A1 case, accuracy will drop by 6.21\% when channel 23 of layer 1 is s@1 leading to an overall worst-case accuracy (min column) of 73.01\%.  
Here, a single channel can be identified that has a significant effect to the overall accuracy of the system.
Even higher accuracy drops can be noticed at higher precisions. As expected, the higher the precision, the more accurate is the error free network. 
On the other hand, CNV-W1A2 shows a worst-case accuracy drop of 10.81\% in layer 2, with an accuracy of 71.85\% which is even lower than the worst-case for W1A1 (73.01\%). This higher drop can be explained by the fact that, virtually, the error injection of s@$-1$ consists of 2 concurrent faults in the coding at bit level. 
However, increasing the weight precision to 2 bits instead achieves, as expected, a slightly better worst-case accuracy of 74.18\% (drop of 10.11\%) compared to CNV-W1A2.

\input{tables/sat_lfc_mix.tex}
\input{tables/sat_channel_mix.tex}


Interestingly, in each network it is possible to identify complete channels which, when fixed to a certain value, actually show higher accuracy (max column) than the baseline. 
As an example, fixing the output channel 2 of layer 0 to value 1 in the W1A1 case would increase the overall accuracy of the network by 0.54\%.
This shows that, despite the quantization, the neural network still contains redundancy, maybe due to overfitting or limitations in training. 
Channel pruning could be applied to those neural networks without loss in accuracy and without the need of retraining but this is out of the scope of this paper.

It can be concluded that QNNs containing convolutional layers are by far not as robust to faults as commonly believed. A single neuron stuck-at at the wrong place can cause accuracy drops of up to 10\% which may not be tolerable in a safety critical application.

\subsection{Robustness Optimization by Selective Channel Replication}
\label{sect:optimization_by_replication}

\input{tables/overhead_channels.tex}
In safety critical applications, fault-tolerant systems rely on the addition of redundancy, e.g., using Triple Modular Redundancy (TMR) to achieve the desired safety integrity level.
Clearly, TMR adds an overhead of 200\%.
To avoid this, we propose selective channel replication as a method to increase the error tolerance by triplicating only the critical channels.
This is performed by first identifying the channels of the neural network that cause accuracy drops which are below a certain threshold. Next, only those channels are replicated using TMR. 

Table~\ref{tab:overhead_channels} lists, for multiple network precision bit widths, how many channels, when affected by a channel s@, would incur a worst-case accuracy drop of more than 0.5\%, 1\% or 2\% in presence of a single error. 
The last row shows the percentage of operational overhead (counted in number of multiply and accumulation operations) in case of triplicating the channels which would ensure a worst-case accuracy drop that is lower or equal to the given thresholds.
As it can be noticed, the lower the tolerated worst-case drop, the higher the amount of overhead, but in all cases the overhead is always smaller than the 200\% given for a full implementation of TMR. 
Giving the results of the error injection, it is possible to evaluate how many channels need TMR in order to achieve a desired single-error worst-case accuracy, leading to minimal increase in the total number of operations. 

\subsection{Cost Analysis of Selective Channel Replication}
In this section, we analyze the trade-off between accuracy and hardware cost across a spectrum of precisions. 
For this, we adopted the hardware cost model for High-Level Synthesis designs for MAC blocks, as explained in~\cite{finn-r}, using the formula $1.6 \cdot w\cdot a$, where $w$ and $a$ are, again, the bit widths for weights and activations, respectively.
The resulting design space is shown in Figure~\ref{fig:pareto}. 
For each precision, the relationship between the hardware cost and the worst-case error rate in presence of a single error is shown. The optimal compromises in the design space will be located on the Pareto frontier.

\input{images/pareto.tex}

Interestingly, each one of the precisions in the CNV networks on CIFAR-10 have entry points into the Pareto optimal curve. 
If the tolerated worst-case error rate is higher than 21.8\%, the completely binarized solution is the optimal solution. 
However, when more channels in the binarized network require triplication to achieve the desired worst-case error rate, then the network with 2 activation bits becomes optimal. 
As expected, for high worst-case accuracy requirements, the solutions with higher precision for both weights and activations provide the best (and only) solution.

%% file: images/satchannel.tex
\begin{figure}
\centering
\resizebox{\columnwidth}{!}{
\begin{tikzpicture}
	\newcommand{\xnor}[2]{
		\filldraw[fill=white, draw=black] (#1, #2) rectangle (#1 + 0.5, #2 + 0.5);
		\draw(#1 + 0.25, #2 + 0.22) node {=};
	};
	
	\foreach \a in {5,...,0}
		\def \x { \a / 10}
		\filldraw[fill=white, draw=black] (0 + \x,0 + \x) rectangle (2 + \x,2 + \x);
		
	\foreach \a in {5,...,0}
		\def \x { \a / 10}
		\filldraw[fill=white, draw=black] (5 + \x,0 + \x) rectangle (6 + \x,2 + \x);
		
	\foreach \a in {5,...,1}
		\def \x { \a / 10}
		\filldraw[fill=white, draw=black] (9 + \x,0 + \x) rectangle (11 + \x,2 + \x);
	\filldraw[fill=white, draw=red, line width = 0.4mm] (9, 0) rectangle (11, 2);
	
	\draw[<->] (-0.05, 2.05) -- (0.45,2.55) node[pos=0.5,sloped,above] {ifm chan};
	
	\draw[<->] (4.95, 2.05) -- (5.45,2.55) node[pos=0.5,sloped,above] {\# PEs};
	
	\draw[<->] (8.95, 2.05) -- (9.45,2.55) node[pos=0.5,sloped,above] {ofm chan};

	\draw(1,1) node {IFMap};
	
	\draw(5.5,1) node {PE};
	
	\draw(10,1) node {OFMap};

	\draw(10,1) node {$ \color{red}
	\begin{matrix}
	1 \, \, \, & \ldots & \, \, \, 1  \\
	\vdots  \, \, \, &  & \, \, \, \vdots  \\
	1  \, \, \, & \ldots & \, \, \, 1 
	\end{matrix}
	$};	
	
	\draw[->, line width=0.4mm] (2.8,1) -- (4.8,1) node[pos=0.5,sloped,above] {input activations};	
	
	\draw[->, line width=0.4mm] (6.8,1) -- (8.8,1) node[pos=0.5,sloped,above] {results};
	
	\draw[->, line width=0.4mm] (5.8,4) -- (5.8,2.8) node[pos=0.5,sloped,above] {weights};

	\draw[dashed] (5,0) -- (1, -0.5);
	\draw[dashed] (6,0) -- (10,-0.5);
	\draw (5.5, -0.5) node {PE in MVTU};
	
	\xnor{1.5}{-1.5}
	\draw (1.75, -1.875) node {\vdots};
	\xnor{1.5}{-3}	
	
	\filldraw[fill=white, draw=black] (2.75, -3) rectangle (4.25,-1);	
	\filldraw[fill=white, draw=black] (5.5, -3) rectangle (7,-1);	
	\filldraw[fill=white, draw=black] (8.25, -3) rectangle (9.75,-1);
	
	\draw (3.5,-2) node[text width = 1cm] {mul};	
	\draw (6.25,-2) node[text width = 1cm] {acc};	
	\draw (9,-2) node[text width = 1.1cm] {thresh \newline olding \newline $th_i$};
	
	\draw (0, -2) node[text width = 1.5 cm] {weights \& input activations};

	\draw[->] (4.25, -2) -- (5.5, -2);
	\draw[->] (7, -2) -- (8.25, -2) node[pos=0.5,sloped,above] {$val$};
	\draw[->, color=red, line width=0.4mm] (9.75, -2) -- (11, -2) node[pos=0.5,sloped,above, color=black] {results};
	
	\draw[->] (2,-1.25) -- (2.75, -1.25);
	\draw[->] (2,-2.75) -- (2.75, -2.75);
	
	\draw[->] (1, -1.125) -- (1.5, -1.125);
	\draw[->] (1, -1.375) -- (1.5, -1.375);
	\draw[->] (1, -2.625) -- (1.5, -2.625);
	\draw[->] (1, -2.875) -- (1.5, -2.875);
	
	\draw[->] (5, -3.5) -- (9, -3) node[pos = 0.0, below] {fixing thresholds for 1 channel};
	
	\draw[->] (10, -3.5) -- (10.5, -2) node[pos = 0.0, below] {stuck-at results for channel};
	
\end{tikzpicture}
}
\caption{Stuck-at injection for a complete channel} \label{fig:satchannel}
\end{figure}
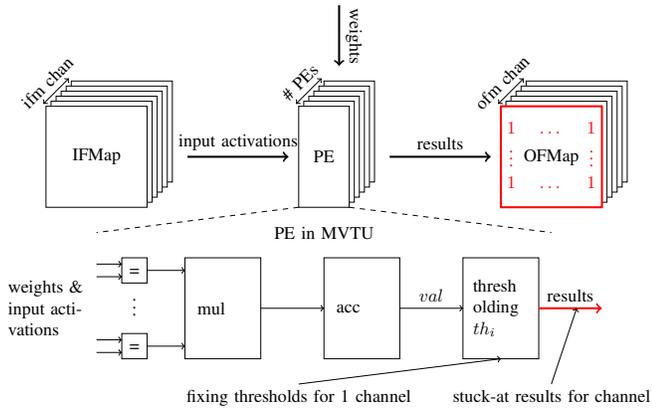

%% file: tables/sat_lfc_mix.tex
\begin{table}[!t]
\caption{Whole channel stuck-at effects on LFC-W1A1, LFC-W1A2 and LFC-W2A2 networks (trained on MNIST)}
\label{tab:satchannel-lfc-summary}
\centering
\begin{tabular}{R{0.5cm}C{0.5cm}R{0.35cm}R{0.35cm}R{0.35cm}R{0.35cm}R{0.35cm}R{0.35cm}}
\toprule
        & \multicolumn{7}{c}{Accuracy in \%} \\
\cmidrule(rl){2-8}
Net & Err. & \multicolumn{2}{c}{single s@$-1$} & \multicolumn{2}{c}{single s@0} & \multicolumn{2}{c}{single s@1}\\
\cmidrule(rl){3-4} \cmidrule(rl){5-6} \cmidrule(rl){7-8}
        & free   & min & max & min & max & min & max\\
\midrule
W1A1 & 98.40 & 98.33 & 98.46 & --    & --    & 98.33 & 98.45\\
W1A2 & 98.49 & 98.44 & 98.55 & 98.45 & 98.54 & 98.43 & 98.55\\
W2A2 & 98.53 & 98.44 & 98.57 & 98.46 & 98.55 & 98.46 & 98.55\\
\bottomrule
\end{tabular}
\end{table}

%% file: tables/sat_channel_mix.tex
\begin{table}[!t]
\caption{Whole channel stuck-at effects on CNV-W1A1, CNV-W1A2 and CNV-W2A2 networks (trained on cifar-10)}\label{tab:satchannel-cnv-summary}
\label{tab:satchannel-cnv-W1A1}
\centering
\begin{tabular}{R{0.5cm}C{0.5cm}R{0.35cm}R{0.35cm}R{0.35cm}R{0.35cm}R{0.35cm}R{0.35cm}}
\toprule
        & \multicolumn{7}{c}{Accuracy in \%} \\
\cmidrule(rl){2-8}
Net & Err. & \multicolumn{2}{c}{single s@$-1$} & \multicolumn{2}{c}{single s@0} & \multicolumn{2}{c}{single s@1}\\
\cmidrule(rl){3-4} \cmidrule(rl){5-6} \cmidrule(rl){7-8}
        & free   & min & max & min & max & min & max\\
\midrule
W1A1 & 79.22 &  75.30 & 79.76 & -- & -- &  73.01 & 79.69\\
W1A2 & 82.66 & 73.81 & 83.24 & 79.91 & 83.18 & 71.85 & 83.11\\
W2A2 & 84.29 & 74.80 & 84.68 & 82.44 & 84.69 & 74.18 & 84.76\\
\bottomrule
\end{tabular}
\end{table}

%% file: tables/overhead_channels.tex
\begin{table*}[!ht]
\caption{Number of channels to be triplicated and operation overhead in order to achieve a certain worst-case accuracy drop}\label{tab:overhead_channels}
\centering
\begin{tabular}{C{0.25cm}ccccccccccccccc}
\toprule
\multicolumn{4}{c}{}  & \multicolumn{3}{c}{CNV-W1A1} & \multicolumn{3}{c}{CNV-W1A2} & \multicolumn{3}{c}{CNV-W2A2} & \multicolumn{3}{c}{CNV-W4A4}   \\ 
\cmidrule(rl){5-7} \cmidrule(rl){8-10} \cmidrule(rl){11-13} \cmidrule(rl){14-16}
& layer &  type &channels & $\geq$ 0.5\% & $\geq$ 1\% & $\geq$ 2\% & $\geq$ 0.5\% & $\geq$	 1\% & $\geq$ 2\% & $\geq$ 0.5\% & $\geq$ 1\% & $\geq$ 2\% & $\geq$ 0.5\% & $\geq$ 1\% & $\geq$ 2\% \\
\midrule
\multirow{8}{0.5cm}{\rotatebox{90}{\#channels to trip.}}
& 0 & conv & 64 & 17 & 7 & 2 & 12 & 8 & 1 & 19 & 10 & 3 & 5 & 0 & 0 \\
& 1 & conv & 64 & 63 & 51 & 24 & 64 & 59 & 38 & 64 & 61 & 38 & 62 & 57 & 33 \\
& 2 & conv & 128 & 106 & 80 & 35 & 121 & 109 & 71 & 120 & 106 & 73 & 114 & 96 & 67 \\
& 3 & conv & 128 & 113 & 75 & 9 & 125 & 116 & 82 & 124 & 115 & 80 & 123 & 116 & 93 \\
& 4 & conv & 256 & 87 & 8 & 0 & 168 & 58 & 3 & 142 & 36 & 1 & 160 & 37 & 3 \\
& 5 & fc & 256 & 0 & 0 & 0 & 0 & 0 & 0 & 0 & 0 & 0 & 0 & 0 & 0 \\
& 6 & fc & 512 & 0 & 0 & 0 & 0 & 0 & 0 & 0 & 0 & 0 & 0 & 0 & 0 \\
& 7 & fc & 512 & 0 & 0 & 0 & 0 & 0 & 0 & 0 & 0 & 0 & 0 & 0 & 0 \\ 
\midrule
 \multicolumn{4}{c}{\parbox{3cm}{ops overhead when trip. req. ch. [\%]}} & 173.47 & 129.70 & 49.87 & 186.24 & 167.65 & 109.49 & 185.24 & 168.86 & 109.36 & 179.63 & 159.60 & 104.96\\
\bottomrule

\end{tabular}
\vspace{-0.5cm}
\end{table*}

%% file: images/pareto.tex
\pgfplotsset{
   every axis/.append style = {
                    label style={font=\small},
                    tick label style={font=\footnotesize} 
                }
}

\begin{figure}
\begin{tikzpicture}[scale=0.95]
\begin{semilogxaxis}[
  grid =major,
  ylabel={100 - Accuracy [\%]},
  xlabel={Hardware Cost [LUT]},
  legend style={draw=none,
  at={(0.5,1.)},
  legend columns=5,
  anchor=south},
  width=\columnwidth,
  height=7 cm,
  legend style={font=\footnotesize}
]
\addlegendimage{only marks, mark=-,blue}
\addlegendimage{only marks, mark=square,teal}
\addlegendimage{only marks, mark=o,black}
\addlegendimage{only marks, mark=triangle,orange}
\addlegendimage{no marks,red}
\addlegendentry{W1A1};
\addlegendentry{W1A2};
\addlegendentry{W2A2};
\addlegendentry{W4A4};
\addlegendentry{Pareto optimal};
\pgfplotstableread[col sep = comma]{images/data/w1a1.csv}\saa;
\addplot +[ blue, sharp plot, mark size=1pt, mark=- ] 
table [y = Error, x = hw_cost] {\saa};
\pgfplotstableread[col sep = comma]{images/data/w1a2.csv}\saa;
\addplot +[ teal, sharp plot, mark size=1pt, mark=square ] 
table [y = Error, x = hw_cost] {\saa};
\pgfplotstableread[col sep = comma]{images/data/w2a2.csv}\saa;
\addplot +[ black, sharp plot, mark size=1pt, mark=o ] 
table [y = Error, x = hw_cost] {\saa};
\pgfplotstableread[col sep = comma]{images/data/w4a4.csv}\saa;
\addplot +[ orange, sharp plot, mark size=1pt, mark=triangle ] 
table [y = Error, x = hw_cost] {\saa};
\pgfplotstableread[col sep = comma]{images/data/pareto.csv}\saa;
\addplot +[ red, sharp plot, no marks ] 
table [y = Error, x = hw_cost] {\saa};
\end{semilogxaxis}
\end{tikzpicture}
\caption{Pareto frontier of single error-tolerant worst-case test error vs. hardware cost}\label{fig:pareto}
\vspace{-0.5cm}
\end{figure}
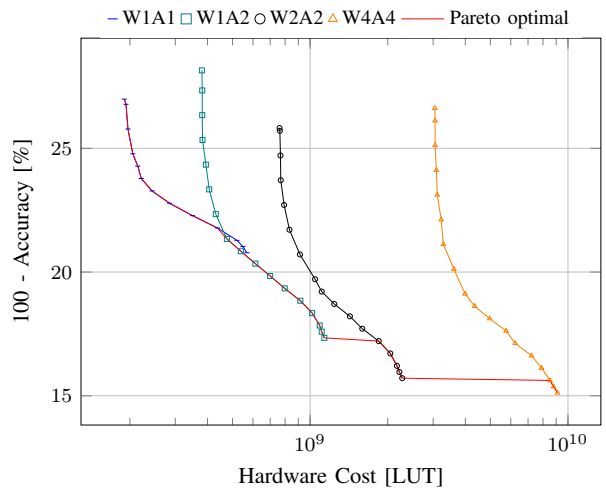

%% file: hardware-dependent.tex
\section{Hardware-Specific Optimization}
\label{sec:hardwaredependent}
Section~\ref{sec:hardwareindependent} showed results of error injection and analyzed the cost of selective triplication. The study was based on the assumption that the neural network accelerator uses parallelism enabling a permanent fault in a PE to affect only a limited portion of a single channel in a single layer. 
Nonetheless, in an overlay architecture, it is more likely that a portion of the same layer or multiple layers are physically executed on the same PE. 
Thus, the presence of a single fault in that PE would affect multiple aspects of the complete neural network, depending on the actual scheduling of the neural network on the array of PEs. 
In this section, we try to explore the effects of a single defective PE evaluating multiple portions of a layer as shown in Figure~\ref{fig:folding}, first by analyzing the scheduling proposed in~\cite{finn}. 

\subsection{Analysis of Stuck-At Errors}
In the default schedule of FINN~\cite{finn}, consecutive channels are computed by consecutive PEs. 
Channel $c$ is computed in PE $=c \mod \#\text{PE}$.
In case of a folding factor $f=\#\text{Channels} / \#\text{PEs}= 2$, with 64 channels and 32 PEs, the first PE will compute channel 0 and channel 32, while the second PE will compute channels 1 and 33 (and so on). 
Figure~\ref{fig:layer0-sa0-PE} shows the accuracy whenever one PE is s@0 in layer 0 of the CNV-W1A1 network and assuming different folding factors for scheduling, listing both the average accuracy among all faulty PEs as well as the difference between the maximum and minimum achieved accuracy. 
As expected, when increasing the folding factor (decreasing the number of PEs) the average accuracy drop increases as a larger portion of the layer result being corrupt by the faulty PE. 
Furthermore, the accuracy drop depends on which specific channels are executed on the faulty PE, making the worst-case accuracy results being schedule-dependent. 
\input{images/folding-error-chart.tex}
It is possible to minimize overall accuracy drop by finding and using the optimal scheduling of channels for a given number of PEs through this analysis as will be analyzed in greater detail below.

\subsection{Robustness Optimization by Scheduling}
As explained above, it is necessary to analyze the effects when combinations of output channels are faulty. To illustrate this, Figure~\ref{fig:scheduling} shows the achieved accuracy when one PE for two output channels (folding factor $f=2$) is s@0, here evaluated for layer 0 of CNV-W1A1. 
For each combination of two channels, which are assumed to be executed on the same PE, the resulting accuracy in case of a single faulty PE is presented, having the first channel on the abscissa and the second channel on the ordinate. 
The worst-case accuracy in case of a single faulty PE with the default scheduling is 74.82\%, indicated with $\diamond$ in Figure~\ref{fig:scheduling}. That would be the case if PE$_{30}$ computing channels 30 and 62 is s@0. 
By analyzing the complete 2-D space in Figure~\ref{fig:scheduling}, it is possible to identify that mapping the calculation of channels 24 and 38 onto the same PE would lead to the even worse single-error accuracy of only 66.86\%.
\input{images/scheduling.tex}
As it is possible to simply reorder the channels, this allows to optimize the schedule for higher error-tolerance. 
To be more precise, it is possible to find the optimal scheduling for providing the most error-tolerant behaviour towards single corrupt PEs.
For that, the following Integer Linear Programming (ILP) model was used to find the schedule that maximizes the accuracy, which is explained in the following:
\begin{align*} 
\text{maximize} \ \text{min}_\text{acc} 
\end{align*}
subject to
\begin{align*}
\text{C1:} && \sum_{j=0,j\neq i}^{N-1} s_{\text{min}(i,j),\text{max}(i,j)} & = 1 \\
           &&            \forall \ i=0\ldots N-1\\
\text{C2:} && (1-s_{ij})M_\text{acc} + s_{i,j}E_{i,j} & \geq \text{min}_\text{acc}\\
           &&            \forall \ i=0\ldots N-1, j=0\ldots i-1\\
           && \text{min}_\text{acc} \in \mathbb{R}, \ s_{i,j}\in\{0,1\}
\end{align*}

The formulation considers a single corrupt PE using the data from the combined error injection analysis and maximizes the minimal accuracy $\min_\text{acc}$ when a PE is corrupt. 
For that, the Boolean variable $s_{i,j}$ is used, which is one, when channel $i$ and $j$ out of $N$ channels are scheduled on the same PE. 
As $s_{i,j}$ is equal to $s_{j,i}$ per definition, we define $i<j$ to avoid redundancy and to achieve a less complex formulation.
Now, any channel $i$ can be scheduled onto a PE with any other channel $j \neq i$, but every channel can only be scheduled once. This is realized by constraint C1. Here min($i,j$) and max($i,j$) are used to achieve $i<j$.
Constraint C2 formulates the minimal accuracy for the scheduling. The constant $E_{i,j}$ is the accuracy, when channel $i$ and $j$ are scheduled on the same PE and that PE is defect. $M_{acc}$ is the highest accuracy the network can achieve when a PE in the layer is corrupt. When the variable $s_{i,j}$ is zero, C2 is fulfilled because $M_\text{acc} \geq \text{min}_\text{acc}$. When $s_{i,j}$ is one, which means that channel $i$ and $j$ are scheduled on the same PE, the constraint becomes $E_{ij} \geq \text{min}_\text{acc}$ which prevents $\text{min}_\text{acc}$ from getting lower than possible. Note that $E_{i,j}$ and $M_\text{acc}$ are the number of correct classifications of the evaluation set with 10.000 images. If we want to maximize the accuracy for s@0 and s@1 together, we can set $E_{i,j}$ to $\text{min}(E_{i,j}^0,E_{i,j}^1)$, where $E_{i,j}^0$ and $E_{i,j}^1$ are the s@0 and s@1 accuracies, respectively. The given formulation is for a folding factor of 2 but can easily be adopted for higher folding factors.

The resulting mapping of channels to PEs of the optimal schedule are marked with a 
$\star$ in Figure~\ref{fig:scheduling}, while the default schedule is marked with a $\diamond$.
The optimal schedule achieves a worst-case accuracy of 76.70\%, which is an increase of 1.88\% with respect to the default scheduling. 
Reordering the scheduling of channels just means rearranging the weights of the network and comes at no additional hardware overhead, making it an appealing solution. 
Nonetheless, there is no easy way to estimate the accuracy for each combination of channels mapped on a faulty PE. 

The analysis relies on time consuming error injection campaigns, in which each possible combination of channels is s@ each possible value.
The number of experiments also heavily increases with the number of channels $c$ and folding factor $f$:
$$\#\text{experiments} = \binom{c}{f} = \frac{c!}{f! \cdot (c -f)!}$$
For example, with $f=4$ and $c=64$, 635,276 experiments are needed for each injected value. 
FPGA hardware acceleration, in case of W1A1 using a Xilinx ZCU104 board (XCZU7EV) running at 300\,MHz, achieves 32,880 frames per second. This implementation enables $\sim$3 experiments per second, finishing the characterization in $\sim$59 hours. This is 5.8$\times$ faster than running Theano using an NVIDIA P40 where the overall campaign took more then 14 days.\\
To conclude, the worst-case accuracy can be further increased by additionally using the Selective Channel Replication method proposed in Section~\ref{sect:optimization_by_replication}. However, when the PEs computing $M$ out of the $N$ channels are replicated, the scheduling should be adjusted to consider only the remaining $N-M$ channels.

%% file: images/folding-error-chart.tex
\begin{figure}
\begin{tikzpicture}
    \begin{axis}[
    ymajorgrids,
    xtick=data,
    xmajorgrids,
    ylabel={Accuracy [\%]},
    xlabel={Folding Factor},
    xmode=log,
    legend style={
    at={(0.4,0.05)},
    legend columns=1,
    anchor=south east},
    log ticks with fixed point,
    log basis x={2},
    ]
    \addlegendimage{red, no marks, line width=1pt}
    \addlegendentry{Baseline};
    \draw [solid, red] (0.0000001,79.22) -- (60,79.22); 
    \addplot [only marks,
    color=black,
    error bars/.cd,
    y dir=both,y explicit, 
    error bar style={color=black},
    ] coordinates {
    (1,78.92828142)     += (0,-3.478284422) -=(0,-0.831720578)
    (2,78.63281306)     += (0,-3.812813063) -=(0,-0.687186937)
    (4,77.85500094)     += (0,-3.015004938) -=(0,-1.314997063)
    (8,76.375)          += (0,-3.605003)    -=(0,-1.724998)
    (16,71.60500125)    += (0,-1.67500125)  -=(0,-0.93499975)
    (32,47.75)          += (0,-11.169998)   -=(0,-11.169998)  
    };
    \addlegendentry{Average}
    \addlegendimage{color=black,mark=|,
    mark options={mark repeat=2,mark phase=1}}
    \addlegendentry{Extrema}    

    \end{axis}
 
\end{tikzpicture}
\caption{Worst-case accuracy with different folding factors on CNV-W1A1 showing layer 0 stuck-at 0 for 64 channels and 64\dots 2 PEs.}\label{fig:layer0-sa0-PE}
\vspace{-0.5cm}
\end{figure}
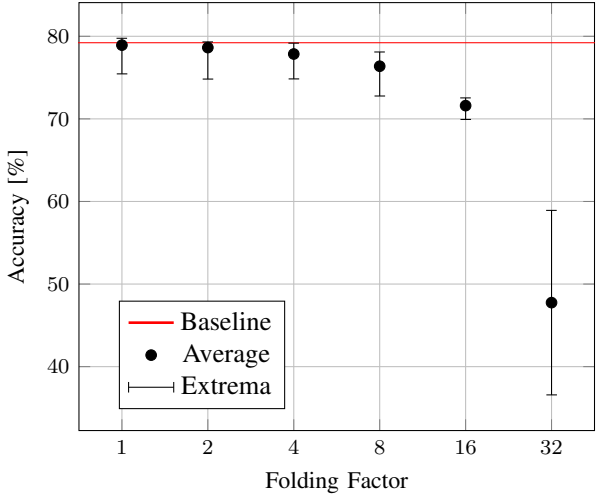

%% file: images/scheduling.tex
\pgfplotsset{
   colormap={cool}{rgb255(0cm)=(255,0,255); rgb255(1cm)=(0,128,255); rgb255(2cm)=(255,255,255)},
   every axis/.append style = {
                    label style={font=\small},
                    tick label style={font=\footnotesize} 
                }
}

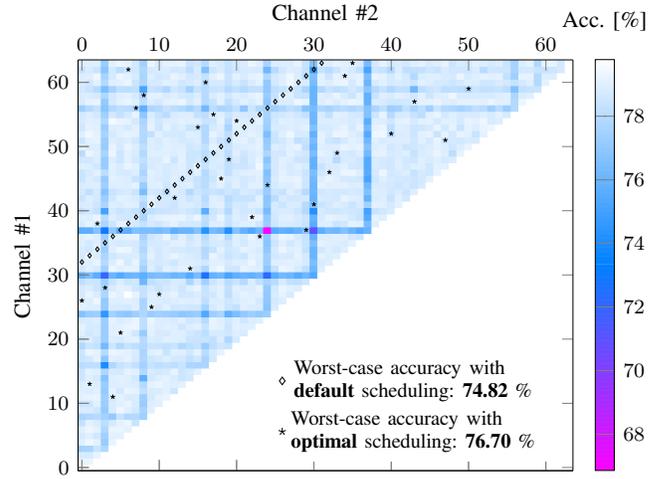
\begin{figure}
\centering
\scalebox{0.96}{
\begin{tikzpicture}
  \begin{axis}[
    view={0}{90},
  	ylabel={Channel \#1},
  	xlabel={Channel \#2},
  	ytick={0,10,20,30,40,50,60},
  	legend style={
        at={(0.95,0.02)},
        draw=none,
        legend columns=1,
        cells={align=left},
        anchor=south east},
    legend style={font=\footnotesize},
    colorbar,
    xticklabel pos=top,
    colorbar style ={
    	  	xlabel={Acc. [\%]},
    	  	xlabel style={at={(0.5,1.14)}},
    	  	width = 7.5pt
    }
  ]
    \addlegendimage{only marks, mark=diamond,black,mark size=1.5pt}
    \addlegendimage{only marks, mark=star,black,mark size=1.5pt}
    \addlegendentry{Worst-case accuracy with \\ \textbf{default} scheduling: \textbf{74.82} \%};
    \addlegendentry{Worst-case accuracy with \\ \textbf{optimal} scheduling: \textbf{76.70} \%};  
    \addplot3[
      surf,
      shader=flat corner,
      mesh/rows=65,
      mesh/cols=65, 
      mesh/ordering=rowwise,
    ] file {images/data/scheduling_new_latex.csv};

    \addplot3[
    only marks,
    mark=diamond,
    mark size = 0.9,
    black
    ] file {images/data/actual_scheduling.csv};

    \addplot3[
    only marks,
    mark=star,
    mark size = 0.9,
    black
    ] file {images/data/optimal_scheduling.csv};
  
  \end{axis}
\end{tikzpicture}
}
\caption{Analysis of different mappings of the computations of two channels to one faulty PE in layer 0 using a folding factor of 2 (32 PEs, 64 channels).}
\vspace{-0.5cm}
\label{fig:scheduling}
\end{figure}

%% file: conclusions.tex
\section{Conclusions}
\label{sec:conclusions}
In this paper, we presented a methodology to characterize and systematically increase the robustness of QNNs with an FPGA-accelerated error injection analysis and two orthogonal methods for robustness optimization. 
The FPGA-accelerated error characterization enables the fast identification of the most susceptible channels. 
Here, our experiments showed that QNNs with convolutional layers are less robust to single faults than commonly believed.
As a countermeasure, the information retrieved from the analysis can be used to decide which channels have to be replicated to achieve a given level of error-tolerance in the generic model. This allows to trade robustness against hardware complexity leading to significant resource reductions compared to common TMR schemes.
In the common case when PEs compute multiple channels, we showed how scheduling influences the robustness and proposed an ILP model that finds the optimal schedule regarding single-fault robustness without any hardware overhead. 
Future work will analyze modeling of hardware errors during training to train error-tolerant neural networks, as well as generalizing the injection tool to model single-event upsets (SEU). 